\documentclass[number]{elsarticle}
\usepackage{amsmath,epsf,amssymb,latexsym,enumerate,shadow,array,color,slashed,ulem}

\DeclareFontFamily{U}{rsf}{}
\DeclareFontShape{U}{rsf}{m}{n}{
  <5> <6> rsfs5 <7> <8> <9> rsfs7 <10-> rsfs10}{}
\DeclareMathAlphabet\Scr{U}{rsf}{m}{n}

\newcommand{\be}{\begin{equation}}
\newcommand{\ee}{\end{equation}}
\newcommand{\ba}{\begin{array}}
\newcommand{\ea}{\end{array}}
\newcommand{\bea}{\begin{eqnarray}}
\newcommand{\eea}{\end{eqnarray}}
\def\IC{\mathbb{C}}

\def\IZ{\mathbb{Z}}

\def\IP{\mathbb{P}}

\def\CM {{\cal M}}

\def\CL {{\cal L}}
\def\CV {{\cal V}}

\def\CK{{\cal K}}

\def\cm{\mbox{cm}}

\newcommand{\eq}[1]{Eq.~(\ref{eq:#1})}

\def\one{{\hbox{ 1\kern-.8mm l}}}

\def\ba{\bar{a}}

\begin{document}

\title{Calabi-Yau metrics and string compactification}

\author[mrd]{Michael R. Douglas}
\ead{mdouglas@scgp.stonybrook.edu} 

\address[mrd]{Simons Center for Geometry and Physics, 
Stony Brook University, Stony Brook NY 11794}

\def\Large{\large}
\def\LARGE{\large\bf}


\begin{abstract}
Yau proved an existence theorem for Ricci-flat K\"ahler metrics in the 1970's,
but we still have no closed form expressions for them.
Nevertheless there are several ways to get approximate expressions, both numerical and analytical.
We survey some of this work and explain how it can be used to obtain physical predictions
from superstring theory.

\noindent
To appear in Nuclear Physics B special issue: 60 Years of  Calabi Conjecture.
\end{abstract}

{\hfuzz=10cm\maketitle}


\section{Introduction}\label{sec1}

Superstring theory is formulated in ten space-time dimensions, and to describe our universe we must postulate that
the six extra spatial dimensions form a compact manifold $\CM$ of diameter smaller than $10^{-17} \cm$.
To a good approximation, the metric on $\CM$ must satisfy the equations of motion of general relativity, which set
the Ricci tensor of $\CM$ equal to a tensor which describes the energy of the vacuum.  In the simplest case, we 
postulate that the vacuum has zero energy, and thus $\CM$ must have a Ricci-flat metric.  

One's first thought might be to take $\CM$ to be the six-torus.  However, this does not work, 
because it leads to a four dimensional theory with too much symmetry to describe our universe.
In particular, every isometry of $\CM$ leads to a particle in four dimensions (a ``graviphoton'') which
can be excluded by observation.  Thus, $\CM$ must be a Ricci-flat manifold without isometries.\footnote{
One might imagine fixing this problem by postulating other fields or structures on $\CM$ which are not invariant 
under the isometry.  While the idea is reasonable, in the end the torus still doesn't work.}

Such manifolds are extremely rare.  In six dimensions, there is only one known class of examples,
the complex K\"ahler manifolds of vanishing first Chern class.  These conditions imply
that the Ricci tensor is globally a total derivative, and that starting from a given K\"ahler potential,
in principle one can add a function to  set the Ricci tensor to zero.  Calabi conjectured in 1957 that this was so,
but to show this one must show that a highly nonlinear
Monge-Ampere equation always has a solution.
Yau famously proved Calabi's conjecture for Ricci flat K\"ahler manifolds in 1977. \cite{Yau}
Such manifolds $\CM$, which are universally known as Calabi-Yau manifolds, are thus the simplest 
candidates for the extra dimensions in our universe.

To get a compactification which solves all of the superstring equations of motion and leads to the Standard Model
gauge group $SU(3)\times SU(2)\times U(1)$ at low energies requires a few more ingredients.  The most important
is that $\CM$ must carry a holomorphic vector bundle $\CV$ of rank 3, 4 or 5, with a connection
which solves the Yang-Mills equations.    This requirement is deduced by identifying the Standard Model gauge
group with the commutant of the holonomy group of $\CV$ in $E_8$ (one of the two $E_8$'s of the
ten-dimensional gauge group).  For the holonomy groups $SU(3)$, $SU(4)$ and $SU(5)$, the commutants are
$E_6$, $SO(10)$ and $SU(5)$ respectively.  These are the ``grand unified gauge groups'' and are in striking
distance of the Standard Model gauge group, in the sense that after tensoring $\CV$ with a nontrivial flat bundle 
(so, with a finite holonomy group), the commutant will indeed be $SU(3)\times SU(2)\times U(1)$.  This
construction favors a non-simply connected $\CM$, so that the flat bundle will be easily obtained.

Much as for the Ricci flat metric, the simplifications of K\"ahler
geometry allow reducing the Yang-Mills equations to a simpler form -- the hermitian Yang-Mills equation --
and lead to a testable necessary condition for a solution, that $\CV$ is $\mu$-stable.
By theorems of Donaldson and Uhlenbeck-Yau, this condition is sufficient.
Anomaly cancellation in  superstring theory constrains the Chern classes of $\CV$ to be
$c_1(\CV)=0$ and $c_2(\CV)=c_2(\CM)$.  While it is not immediately obvious which $\CM$ carry such bundles,
for $\CM$ Calabi-Yau, there is an evident candidate: $\CV\cong T\CM$, the tangent bundle of $\CM$.

These ingredients were first put together in 1985 in a famous paper of Candelas, Horowitz, Strominger and Witten
(CHSW). \cite{CHSW}
This was the first proposal ever for a derivation starting from a fundamental physical theory which could lead
all the way to the Standard Model.  Since then, while other proposals have been made, 
this remains the prototypical and arguably the best proposal yet made for
the fundamental structure which leads to the physical laws of our universe.  The study of Calabi-Yau
compactification also led to the
discovery of mirror symmetry, and many other developments in the interface between string theory and mathematics.

After the gauge group, the next real world observable to reproduce is the number of
generations of quarks and leptons, which is three in the Standard Model.  In CHSW this arises as the index of a Dirac operator
acting on sections of $\CV\otimes \Lambda^2 T\CM$.  Following the ansatz that $\CV\cong T\CM$,
this index can be shown to equal half the Euler character of $\CM$, and thus one wants a Calabi-Yau
manifold with $\chi=6$.   One furthermore wants $\pi_1(\CM)\cong \IZ_3\times\IZ_3$ to get a flat bundle breaking
$E_6$ to the Standard Model gauge group.
And indeed, such a manifold was very quickly constructed \cite{Yau2,TianYau}
and used for string phenomenology. \cite{Greene}
After thirty years, although there are a few competitors \cite{Candelas:2007ac}, this
``three generation manifold'' is still the leading candidate within this class of model.

Over the years, several other classes of string compactifications have been found that can reproduce the
Standard Model.  These include the heterotic ``$(0,2)$'' models, 
type II orientifold compactifications
with Dirichlet branes, F theory on elliptically fibered Calabi-Yau fourfolds, and M theory on manifolds of
$G_2$ holonomy.  If we assume that there is low energy supersymmetry (meaning low compared to the
compactification scale), this may be the complete list.\footnote{
The main exception to this claim are the ``non-geometric'' compactifications which
rely for their existence on special features of string theory, such as nonperturbative effects on the world-sheet 
or in space-time, or non-geometric monodromies.  On the other hand, there are many cases in which geometric
interpretations for such constructions were found later, so the claim has not yet been disproven.}

The geometric heterotic $(0,2)$ models, also called ``superstrings with torsion,''
are obtained by taking $\CV$ different from $T\CM$, though
still satisfying the anomaly cancellation constraints.  As an example, one could take 
a deformation of $T\CM\oplus \IC^K$.
The metric and gauge connection now must satisfy
a set of equations given by Strominger in \cite{Strominger:1986uh}, involving an additional scalar field.
In \cite{Li:2004hx} Li and Yau proved that solutions exist for small
deformations of $T\CM\oplus \IC^K$.\footnote{
This was generalized to certain non-K\"ahler manifolds in \cite{Fu:2006vj}.}
The number of generations in a $(0,2)$ model is half the third
Chern class of $\CV$, opening a new way to get three generations.  
Many explicit examples are known, such as \cite{Braun:2011ni}
which obtains precisely the spectrum of the minimal supersymmetric Standard Model.

\section{Physical predictions from Calabi-Yau manifolds}

Once one has matched the particle content of the Standard Model, one would like to make more
detailed physical predictions, such as the masses of quarks and leptons.  Needless to say,
this is a long story, but for the well-understood constructions it can be summarized as follows.

First, in the present state of the art, quantum corrections must be
small at the compactification scale.  This is simply because we do not have a complete formulation
of string/M theory which could be used to make general computations with quantum
corrections.  Certainly, the use of duality arguments has led to major advances in this direction, but they do not 
yet reach the goal.  Thus, our calculations will be based on the geometry of metrics and connections.

Related to this, we need to assume $N=1$ supersymmetry at low energies.  There are
many arguments for this, but the simplest is that supersymmetry drastically constrains
the quantum corrections and forces many of them to be zero.  In this case, the masses of quarks
and leptons are determined by the ``cubic Yukawa couplings.''  These are the amplitudes for a
given quark or lepton to interact with the Higgs field, which after electroweak symmetry breaking
gives mass to the quarks and leptons.  Given supersymmetry, the Higgs field has a fermionic
partner (the ``Higgsino'') which is not fundamentally different from the quarks and leptons; thus these couplings
have complete permutation symmetry.  There are 
many further details: for example supersymmetry requires there to be two Higgs superfields, there
are logarithmic corrections from the renormalization group, and so on, but the basic inputs are
the cubic Yukawa couplings.

Thus, a good approximation to the problem for the heterotic string is the following.  We start with ten-dimensional
super Yang-Mills theory, and find explicit normalized zero modes of the Dirac operator on $\CM$ coupled to the
Yang-Mills connection on $\CV$, label these $\psi_{a}$.  We then compute the triple overlap of these wave functions,
\be
C_{abc} = \int_\CM \psi_{a} \cdot \psi_{b} \cdot \psi_{c} .
\ee
The case with $a$ and $b$ corresponding to a particular quark, and $c$ corresponding to the Higgsino
determines the ratio of the quark mass to the Higgs expectation value.  For example, for $a=b$ the top
quark and $c$ the ``up'' Higgs, matching the observed top quark mass requires $C = 1$ to a fairly good
accuracy (!)

One can go some distance in this direction without knowing the metric on $\CM$, because the
zero modes have a topological interpretation. \cite{Strominger:1985ks}
The best case is for $\CV\cong T\CM$, as then
K\"ahler geometry relates the zero modes of the Dirac operator to harmonic forms.  On a Calabi-Yau
threefold, the nontrivial harmonic forms will be of Hodge type $(1,1)$ and $(2,1)$, and Hodge duals of
these.  These correspond to ``generations'' and ``antigenerations'' of quarks and leptons respectively,
with gauge charges in the fundamental and antifundamental of $E_6$.  For a more general $\CV$,
the relevant forms are the $(0,p)$-forms with values in $\CV$ and its various tensor power bundles.

One can even get exact formulas for some Yukawa couplings, such as those between a triplet of $(2,1)$-forms.
These are
\be\label{eq:holo}
C_{abc} = \int_\CM \Omega^{3,0} \wedge \left[ \psi_{a} \wedge \psi_{b} \wedge \psi_{c} \right],
\ee
where $\Omega^{3,0}$ is the holomorphic three-form, and the brackets denote contractions of the
holomorphic indices of the $\psi$'s with further $\Omega^{3,0}$'s.  Importantly, this formula is 
``topological'' in that it does not depend on the specific $(2,1)$-forms we use, only their
cohomology class, and thus we do not need to know the metric on $\CM$ to compute it. 
Even better, it can be shown to receive no string world-sheet corrections.
Many examples have been worked out, {\it e.g.} see  \cite{Anderson:2009ge}.

While \eq{holo} is very powerful, it is not the answer to the physical question
we asked unless we can compute it for {\it normalized} $(2,1)$-forms.
The straightforward way to do this is to do the integrals
\be \label{eq:kahler}
K_{a,\bar b} = \int_\CM \omega\wedge\omega\wedge \left[ \psi_{a} \wedge \bar\psi_{\bar b} \right]
\ee
(where $\omega$ is the K\"ahler form)
and go to a basis in which $K_{a,\bar b}=\delta_{a,\bar b}$.  This formula does depend on the 
specific representatives we take for the $\psi$'s, and finding the harmonic forms 
requires knowing the Ricci-flat metric on $\CM$.

In some cases, there are ways around this.   In particular, when $\CV\cong T\CM$,
one can compute \eq{kahler} using ``special geometry.''  Mathematically, deformations
of $T\CM$ integrate to deformations of the complex moduli of $\CM$, and the metric on
this moduli space
can be derived using Weil-Petersson geometry.  However,
the generalization of this approach to other $\CV$ is not known.  Thus, the next step in obtaining
physical predictions is to find the Ricci-flat metric on $\CM$, the hermitian Yang-Mills connection
on $\CV$, and the normalized $\CV$-valued $(0,p)$ forms.

\section{Analytic descriptions of Ricci-flat metrics}

At present there are no closed form expressions for the Ricci-flat metric on any
nontrivial compact Calabi-Yau.  This includes the K3 manifold which is even hyperk\"ahler
and (in principle) amenable to the twistor transform, as well as moduli spaces of instantons
on K3 which are (again in principle) amenable to the Nahm transform.  This is not for want
of trying and thus we can confidently state that this is a ``hard problem.''

There are a number of explicit expressions for Ricci-flat metrics on noncompact Calabi-Yau
manifolds, starting with the ``self-dual gravitational instantons'' of Eguchi-Hanson and the
Taub-NUT spaces.  The best cases are the resolutions of the canonical singularities $\IC^2/\Gamma$
with $\Gamma$ a discrete subgroup of $SU(2)$.  These are hyperk\"ahler and there are explicit
hyperk\"ahler quotient constructions for the metrics.  In particular, it is easy to see that these
metrics are asymptotically locally Euclidean (ALE).  One can also construct moduli spaces
of Yang-Mills instantons on these resolved quotients with
their hyperk\"ahler metrics. \cite{Kronheimer-Nakajima}

As was pointed out by Page \cite{Page}, one can get a nice picture of certain Ricci-flat K3 metrics by
starting with the flat quotient $T^4/\IZ_2$, excising the neighborhood of the fixed points, and gluing
in an Eguchi-Hanson metric at each one.  These are generally known as Kummer surfaces.
This idea has been much used in existence proofs, 
for example in Joyce's construction of metrics of $G_2$ holonomy. \cite{Joyce}
Using these techniques,
in principle one could patch together harmonic forms to compute \eq{kahler} for resolved orbifolds.

The other general construction of explicit Ricci-flat metrics is to assume so much continuous
symmetry that the K\"ahler potential becomes a function of a single variable.  The Monge-Amp\`ere
equation then reduces to an ODE.  This was done for $\IC^n/\IZ_n$ by Calabi. \cite{Calabi}  
There are many examples in the physics literature, where they are known as ``cohomogeneity one'' metrics.

Once one moves on to $\IC^3/\Gamma$,
or other K\"ahler manifolds, while many cases still admit quotient constructions, 
these are symplectic quotients which have no reason to be Ricci-flat.  An amusing case for which
one can get a Ricci-flat threefold from a quotient is to realize $\IC^3/\IZ_3$ using the McKay quiver,
and boldly continue the two moment map 
(Fayet-Iliopoulos) parameters to complex conjugate values.  This procedure reproduces
Calabi's metric, without the need to solve any differential equations. \cite{Douglas:1997zj}
I wonder if this has some deep significance.

The Strominger-Yau-Zaslow approach to mirror symmetry \cite{Strominger:1996it}
suggests the following strategy.  One starts with a torus fibration $T^n\rightarrow \CM\rightarrow B_n$
and a ``semi-flat structure'' on $\CM$.  This is a map from $B_n$ to the moduli space of flat tori which
can be lifted to a Ricci-flat metric, which however is singular for nontrivial fibrations.
One then adds a series of instanton corrections which smooth out the singularities and (again in 
principle) sum to a smooth Ricci-flat metric.
An interesting and related recent development is the work of Gaiotto, Moore and Neitzke
\cite{Gaiotto:2008cd}
on hyperk\"ahler metrics and instanton corrections in four-dimensional supersymmetric gauge theory.

\section{Numerical descriptions of Calabi-Yau metrics}

Even if analytic expressions for Calabi-Yau metrics are found someday, it seems likely
that they will be very complicated.  This is a particularly safe bet for realistic 
string compactification manifolds and bundles,
which are not so simple even from the algebraic geometric point of view.  

For many purposes, not just our string compactification questions,
it would be more useful to have a simple rough description of the metric, which could be used to compute
its properties to order one (but controlled) accuracy.  Numerical methods would seem well suited for this goal.

The numerical study of Calabi-Yau metrics was initiated by Headrick and Wiseman \cite{Headrick:2005ch} who studied 
Kummer surfaces.  Their method was to discretize the Monge-Ampere equation on an explicit lattice 
obtained by introducing coordinate patches, one on the $T^4/\IZ_2$ quotient and one on the Eguchi-Hanson
patch (by symmetry one can force all the Eguchi-Hanson patches to be isometric).  They then applied
Gauss-Seidel relaxation to solve this equation.  Besides exhibiting the corrections to the flat and
Eguchi-Hanson patch geometries, they went on to get a low-lying eigenvalue of the scalar Laplacian.

While the lattice leads to fairly good results, accurate to around 1\% with a few days work on a desktop
computer, it comes with some problems.  In $D$ dimensions, a lattice with linear extent $N$ has
$N^D$ lattice sites, so for $D=6$ a computer memory will only allow $N\sim 20$ or so.  Numerical
accuracy is generally proportional to $1/N$.  While 5\% global accuracy is better than we might need, the
curvature is usually concentrated in certain regions, and this forces us to larger $N$.
This sort of problem is usually dealt with using adaptive or multiscale methods, in which the lattice
spacing varies according to the local gradients of the functions involved.  An adaptive discretization
scheme on a manifold of complicated topology would be rather complicated to implement.

Even using a simpler discretization, a good deal of work is required to
find explicit coordinate patches and overlap functions.  It would be much better if the discretization could
be derived from the geometry of the manifold, either intrinsically or using some simple embedding.
This was done in the work of Donaldson \cite{Donaldson}, which introduced several new ingredients.  

First, a natural finite dimensional
space of K\"ahler metrics can be obtained by embedding $\CM$ using sections of a line bundle $\CL$, and
using as metric the curvature of a unitary connection on $\CL$.  As Donaldson comments, 
``The potential utility of this point of view has been advocated over many years by Yau,''
and indeed it turns out to be quite useful here.

This is much more concrete than it may sound.
Consider a quintic hypersurface in $\IP^4$; what we are doing is looking for the best approximation to
a Ricci-flat metric in the space $\CK_N$ of K\"ahler potentials
\be \label{eq:def-KN}
K_{N,h} = \log \sum_{I_1\ldots I_N,J_1\ldots J_N} H_{I_1\ldots I_N,J_1\ldots J_N} 
Z^{I_1}\ldots Z^{I_N} {\bar Z}^{J_1}\ldots {\bar Z}^{J_N} 
\ee
where $H$ is a constant hermitian matrix.
While one might have thought that this $N$ is no better than the linear lattice extent $N$, in fact the potential
accuracy is better than $N^{-\nu}$ for {\it any} $\nu$ !
This is because the Ricci-flat metrics we are looking for are analytic, and this
behavior is analogous to the fast decay
of Fourier coefficients of analytic functions.

Donaldson then replaced the problem of finding the Ricci-flat metric by that of finding the ``balanced metric.''
We can define this in terms of the following map from hermitian metrics $h$ on $\CL$ to
$\CK_N$.  Given a metric $h$, we compute the integral over $\CM$ of the inner product between sections,
to get a hermitian metric $H^{-1}$ on the space of sections.  We then invert this to get $H$ in \eq{def-KN}.
Applying this map to the hermitian metric $h=e^{-K}$ corresponding to $K$,
we get a map from $\CK_N$ to itself.  The balanced metrics are then the fixed points of this map.
A variant is to assume a given volume form $\nu$ on $\CM$; this type of balanced map exists by
results of Bourguignon, Li and Yau \cite{BLY}.  
For a Calabi-Yau manifold, one can take $\nu=\Omega \wedge\bar \Omega$.

Using the Tian-Yau-Zelditch-Lu expansion for the diagonal of the Szeg\"o kernel,
one can then show that
as $N\rightarrow\infty$, the balanced metric converges to the Ricci-flat metric,
with corrections of order $1/N^2$ (granting $c_1(\CM)=0$).  While we don't have space to repeat these arguments here,
let me mention the paper \cite{Douglas:2008pz} which rederives this expansion using physics techniques.

Donaldson went on to implement this procedure on K3, getting 1\% accuracy now with $N=9$.
While much simpler, one difficult point remained -- namely, the use of explicit coordinates to do integrals over $\CM$.  
Now multidimensional numerical integrals are usually best done by Monte Carlo, and
as it happens it is easy to produce random points drawn from the restriction of a Fubini-Study
measure to a subvariety.  This procedure was implemented in \cite{Douglas:2006rr}, leading to metrics on quintic threefolds
of comparable quality, but with minimal programming effort and computer resources.  The method was used to
study eigenfunctions of the scalar Laplacian in \cite{Braun:2008jp}, and a more accurate
adaptive version of the method was developed in \cite{Anderson:2011ed}.

As another option,
rather than find the balanced metric, one can instead minimize an error term related to the Ricci curvature.
Numerical optimization is quite efficient and this procedure was successfully carried out for a 
Calabi-Yau metric in \cite{Headrick:2009jz}.
See also \cite{Doran:2007zn} and \cite{Bunch:2008}, which studied K\"ahler-Einstein metrics on toric
surfaces using a variety of numerical techniques,

Donaldson's approach can also be used to obtain approximate hermitian Yang-Mills connections
and normalized $(0,p)$-forms.  The idea is to embed $\CV$ by sections into a higher dimensional Grassmannian
manifold $\mbox{Gr}$, and pull back a simple connection on $\mbox{Gr}$ to get a connection on $\CV$
which depends on the parameters of the embedding.  Actually, the anomaly cancellation condition $c_1(\CV)=0$
implies that $\CV$ will have no global sections, but this can be finessed by taking sections of $\CV\otimes\CL^N$ for
some positive line bundle $\CL$.

The concept of balanced metric was defined in this context by Wang \cite{Wang}, who proved
that a series of balanced metrics will converge to the metric associated to a hermitian Yang-Mills connection
as $N\rightarrow\infty$.\footnote{Strictly speaking, one gets a ``weak hermitian Yang-Mills connection''
solving the equation $\omega\cdot F = R(\omega)$  \cite{Wang}.}  
The K\"ahler form $\omega$ on $\CM$, which appears in the
hermitian Yang-Mills equation, is the curvature of $\CL$.  Furthermore, one can adapt the same iterative
procedure to find the balanced metric. If it exists, this procedure was proven to converge by Seyyedali \cite{Seyyedali:2008}.
These ideas were implemented and numerical
solutions found in \cite{Douglas:2006hz,Anderson:2010ke,Anderson:2011ed}.

Now one has a necessary condition for a hermitian Yang-Mills solution, namely that $\CV$ is $\mu$-stable.
If it is not, then the theorem of Wang implies that no balanced metric can exist, and therefore the iterative procedure
cannot converge.   
In \cite{Anderson:2010ke,Anderson:2011ed},
this idea was developed into a numerical method for checking stability. 

To summarize, we now have all of the ingredients required to compute both \eq{holo} and \eq{kahler}.
Although we still do not know which Calabi-Yau manifold makes up the extra dimensions of our universe, 
if and when we do, we will be ready to derive masses and coupling constants from string theory.

\end{document}